\documentclass[prl,amsmath,twocolumn,superscriptaddress,floatfix,letterpaper,notitlepage,noshowpacs]{revtex4-2}
\usepackage[normalem]{ulem}
\usepackage{soul}
\usepackage{epsfig}
\usepackage{amsfonts,amsmath} 
\usepackage{nicefrac}
\usepackage[dvipsnames]{xcolor}
\usepackage[colorlinks=true,citecolor=MidnightBlue,linkcolor=MidnightBlue,urlcolor=MidnightBlue]{hyperref}
\setcitestyle{square}
\usepackage{subfigure} 
\usepackage{natbib} 
\usepackage{graphicx,amssymb} 
\usepackage{tcolorbox} 
\usepackage{tikz}
\usetikzlibrary{shapes,positioning,arrows,calc}

\usepackage{sidecap}
\usepackage{mhchem}
\usepackage[toc,page]{appendix}
\sidecaptionvpos{figure}{t}

\newcommand{\Hquad}{\hspace{0.5em}} 
\newcommand{\HHquad}{\hspace{0.25em}} 
\usepackage{mathrsfs}
\usepackage{commath}

\newcommand{\approptoinn}[2]{\mathrel{\vcenter{
  \offinterlineskip\halign{\hfil$##$\cr
    #1\propto\cr\noalign{\kern2pt}#1\sim\cr\noalign{\kern-2pt}}}}}

\usepackage{gensymb} 
\usepackage{mathtools} 
\usepackage[makeroom]{cancel} 

\usepackage[makeroom]{cancel}

\usepackage{enumitem}

\usepackage{lettrine}
\usepackage{setspace}

\begin{document}
\title{Harvesting Reshapes Dynamical Populations}

\author{R. K. Singh}
\email[]{rksinghmp@gmail.com}
\affiliation{Department of Physics,
Banaras Hindu University,
Varanasi 221005, India
}
\affiliation{Department of Physics,\
  University of Louisiana at Lafayette,\
  Lafayette, LA 70503, USA
} 
\author{Michael~Assaf} 
\affiliation{Racah Institute of Physics,\ The Hebrew University of Jerusalem,\ Jerusalem 9190401, Israel
} 
\author{Jason~R.~Green}
\affiliation{Department of Chemistry,\
 University of Massachusetts Boston,\
 Boston, MA 02125, USA
}
\affiliation{Department of Physics,\
  University of Massachusetts Boston,\
  Boston, MA 02125, USA
} 
\author{Erez~Aghion} 
\email[]{Erez.Aghion@louisiana.edu}
\affiliation{Department of Physics,\
  University of Louisiana at Lafayette,\
  Lafayette, LA 70503, USA
} 

\date{\today}

\begin{abstract}
Harvesting--the periodic removal of individuals above or below a threshold trait value--reshapes heterogeneous populations without altering their underlying stochastic dynamics. We study how repeated harvesting events steer the evolution of probability densities for classes of stochastic processes exhibiting both normal and anomalous dynamics, as well as a prototypical predator-prey model.
Removal of the upper portion of the density drives the system to a quasi-steady state when viewed at the ``harvesting clock''. This state depends only on the harvesting threshold and frequency but not on the initial conditions.
Removal of the lower portion of the density fixes its shape while generating a constant effective drift that exceeds that of the unharvested mean.
Our results suggest the possibility of manipulating the dynamics of stochastic populations through external selection interventions.
\end{abstract}

\maketitle

\newcommand{\fr}{\frac}
\newcommand{\tl}{\tilde}
\newcommand{\lr}{\langle}
\newcommand{\rl}{\rangle}

One of the key drivers of evolution is competition, which leads to natural selection. 
Human interventions such as hunting and fishing can lead to \textit{unnatural selection}~\cite{thambithurai2024environmental, charmantier2024does} by selectively removing individuals with particularly desired or undesired traits. 
For example, larger fish  are typically more susceptible to terminal fishing than smaller ones, which can drive an evolutionary shift toward smaller body sizes~\cite{darimont2009human}. 
Such interventions clip and reshape the distribution of traits in heterogeneous populations. 
Even if the distribution later recovers, it does so from a modified state, thereby altering its subsequent evolution. 
This raises natural questions: How do repeated harvests affect the long-term evolution of a population? 
Can harvesting be deliberately used to steer the future dynamics of a system?

Harvesting is not limited to biological populations. 
Any dynamical system with intrinsic randomness---arising from noise, stochastic parameters, or random initial conditions---can, in principle, be harvested. 
As a simple illustration, consider an ensemble of Brownian particles in contact with a heat bath at constant temperature. 
Starting from a localized initial condition, $\rho_0(x)\!=\!\delta(x)$, the spatial density evolves diffusively as
$
\rho_0(x)\rightarrow \rho_t(x)=(2\pi\sigma^2(t))^{-1/2}\exp\!\left\{-x^2/[2\sigma^2(t)]\right\},
$
with variance $\sigma^2(t)=2Dt$, and diffusion coefficient $D\!>\!0$. 
What will be the shape of the spatial density if, at regular time intervals $\tau,\,2\tau,\ldots,(N-1)\tau$, with $N>1$, we remove all the Brownian particles above or below a fixed percentile of the distribution and focus only on the spatial statistics of the remaining population? Many stochastic systems  experience random fluctuations described  as Brownian motion~\cite{metzler2000random,klafter2011first}. Answering the question may allow  to effectively steer the dynamics of such systems. 

Here, we study the effects of harvesting-induced reshaping of dynamical populations in a range of prototypical stochastic processes: Brownian motion, continuous-time random walk, run and tumble and L\'evy walk, as well as in a predator-prey model. 
We are interested in dynamical changes occurring as a result of the effect of harvesting on the population's \textit{composition}, rather than its size. We show that the system's response to reshaping depends qualitatively on whether the distribution is clipped above or below a fixed threshold value or percentile, suggesting that harvesting can steer the temporal behavior of dynamical populations. 
Noisy dynamical systems are often characterized by a variance that grows as $\sigma^2(t)\sim t^\alpha$, corresponding to normal dynamics for $\alpha=1$ and anomalous dynamics otherwise. 
The effects of periodic reshaping that occur in both types of processes indicate their broad applicability.

\section{Quantitative Analysis}
\vspace{-0.3cm}
\subsection{Mathematical description of unharvested systems}
\vspace{-0.3cm}
To quantify the act of stochastic reshaping of probability densities by harvesting, we first recall the standard mathematical treatment of unharvested systems. Consider a stochastic process on the semi-infinite line $[0,\infty)$, with a reflecting wall at $x = 0$. In what follows, we generally refer to the element undergoing the stochastic process as a ``random walker'', and $x(t)$ is its ``position'' at time $t$. However, as we show, the units and meaning of $x$ can vary depending on the stochastic process governing the system at hand. For an ensemble of random walkers starting
from an initial spatial probability density $\rho_0(x)$, the position density at time $t$ is determined by 
\begin{align}
\rho(x,t) = \int_0^\infty dy~G(x,t|y,0)\rho_0(y). 
\label{EqGreen}
\end{align}
Here, $G(x,t|y,0)$ is the transition probability from $y\rightarrow x$ at a time interval of duration $t$. 
Without external intervention, evolving the system at a series of time intervals $t_1,t_2,\dots t_N$, the initial state transforms from $\rho_0(x) \rightarrow$ $\int_0^\infty {dy_{N-1}}~G\left(x,\sum_{i=1}^N t_i\middle|y_{N-1},\sum_{i=1}^{N-1} t_i\right)\dots$ $\int_{0}^{\infty} {dy_1}~G(y_2,t_2
+t_1|y_1,t_1)\int_{0}^{\infty}{dy_0}~G(y_1,t_1|y_0,0)\rho_0(y_0)$, which is indistinguishable from Eq.~\eqref{EqGreen} with $t=\sum_{i=1}^N t_i$. 

\begin{figure}
    \centering
    \hspace{-0.5cm}\includegraphics[width=0.48\textwidth]{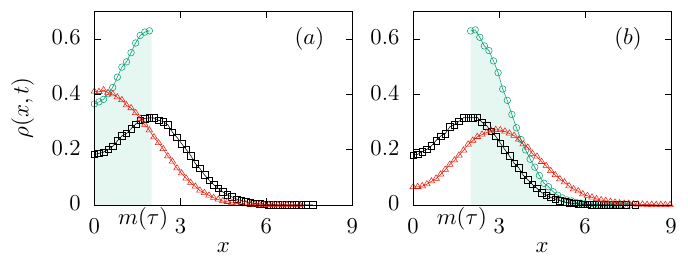}
    \vspace{-0.5cm}\caption{\footnotesize{\textit{Position probability density $\rho(x,t)$ of Brownian particles before and after a harvest at $t=\tau$.} Particles are initially localized at $x=2$. Without harvesting, $\rho(x,t)$ spreads diffusively with a reflecting boundary at $x=0$. Black squares show the density at the instant right before the harvesting, at time $\tau$. Here, the diffusion coefficient is $D=1,\tau=1$. Green circles show the density immediately after reshaping, at $\tau^+$. Red triangles show the density after an additional interval $\tau$.
(a) Backward reshaping: clipping the density above the median $m(\tau)$ and renormalizing below it hinders the subsequent spreading of the density. 
(b) Forward reshaping: removing density below $m(\tau)$ and renormalizing it above this threshold enhances spreading, yielding a broader density farther from the wall.
The median $m(\tau)$ marks the boundary of the green region.
}}
    \label{fig1}
\end{figure}

\subsection{Defining a \textit{reshaping} event in harvested systems}

When a system is harvested, a portion of the population above (below) a threshold value $x_c$ is removed. Here, we assume that this removal occurs over a very short time; thus it is considered as ``instantaneous''. Reshaping is the effect of harvesting on the probability distribution. The harvesting sets the reshaping event transforming $\rho(x,t)$ at time $t = \tau$ to $\rho(x,\tau^+)$: 
\begin{align} 
&\hspace{-0.3cm}\text{Reshaping backward:}\Hquad  \rho(x,\tau^+)=\frac{\rho(x,\tau)\Theta(x<x_c)}{Z}. \nonumber\\ 
&\hspace{-0.3cm}\text{Reshaping forward:}\quad\HHquad \rho(x,\tau^+)=\frac{\rho(x,\tau)\Theta(x>x_c)}{Z},
\label{EqReshaping}
\end{align}  
where $\Theta(\cdot)\equiv 1$ if the condition inside the brackets is satisfied and zero otherwise, and $\tau^+$ is the immediate instant after the reshaping event. Reshaping backward means removing all the population above the threshold, namely this reshaping is performed against the expansion of the probability density (given a reflecting boundary at $x=0$). This process is mathematically related to conditioning a killed Markov process on survival at successive times~\cite{Sylvie2012Quasi}. Forward reshaping is performed in the direction of the natural expansion of the density by removing all the population below the threshold, see Fig.~\ref{fig1},  and is reminiscent of the  Fleming-Viot particle system~\cite{burdzy2000fleming}. 
After clipping the density, we renormalize it using the coefficient $Z$; for backward reshaping; $Z=\int_{0}^{x_c}\rho(x,\tau)dx$, while in forward reshaping $Z=\int_{x_c}^{\infty}\rho(x,\tau)dx$. Through this renormalization, we are able to focus on the shape dynamics of the surviving population, assuming that even after the removal of elements from the ensemble its size is still sufficiently large for producing a smooth probability density. 

Figure~\ref{fig1} illustrates the effect of a single harvesting event on the spatial density of Brownian particles, diffusing from an initial location on the positive $x$-axis, in contact with a constant temperature heat bath and a reflecting wall at $x=0$ (see Appendix~\ref{appendix:reshape} and \ref{appendix:brwn} for simulation details).  The shape of the spatial density prior to harvesting is shown with black circles both in panel (a) and (b) before and after the distribution is reshaped at some time $t=\tau>0$. In panel (a), the distribution is \textit{reshaped backwards}, by removing all individuals above the median and renormalizing the remaining density below it (shown in green). Panel (b) shows the complementary \textit{forward reshaping} protocol, where individuals below the median are removed. In both cases, red triangles indicate the distribution after subsequent evolution over an additional interval $\tau$. Backward-reshaping shifts the density toward smaller $x$-values, slowing the expansion of the density, while forward reshaping accelerates it. 

\subsection{Consecutive stochastic reshapings}

A series of consecutive reshaping events is defined as follows: Consider a probability density function evolving at a series of time intervals $t_1,t_2,\dots t_N$, and harvested to reshape it backward at the end of each interval. These reshaping times set the ``ticks'' of a \textit{harvesting clock}. Starting from an initial state $\rho_0(x)$, at time $t=\sum_{i=1}^N t_i$: 
\begin{eqnarray}
\hspace{-0.2cm}\rho(x,t) =\!\! 
&\int_{0}^{\infty}\!\!dy_{N-1}\, G\left(\!x,\sum_{i=1}^N t_i\middle|y_{N-1},\sum_{i=1}^{N-1} t_i\!\right)\!\frac{\Theta(x<x_c^{(N)})}{Z_N}  \nonumber\\
&\hspace{-1.1cm}\times\cdots\times \int_{0}^{\infty} dy_1\, G(y_2,t_1+t_2|y_1,t_1)\Theta(y_2<x_c^{(2)}) \nonumber\\
&\hspace{-1.7cm}\times \int_{0}^{\infty} dy_0\, G(y_1,t_1|y_0,0)\Theta(y_1<x_c^{(1)})\rho_0(y_0).
\label{EqConsecutivePartialConvolutions}\end{eqnarray}
Here $x_c^{(i)}$ is the threshold after step $i$, and  $Z_N=\int_{0}^{\infty}\rho(x,t)dx$. 
A recursive equation provides the probability density function after the $i$th reshaping event, provided its shape after $i-1$ reshapes, namely at time $t_{i-1}=t_i-\tau$. 
In the case of backward reshaping, we have:
\begin{equation}
    \hspace{-0.2cm}\rho(x,t^+_i) = \frac{1}{Z_{t_i^+}}
 \int_{0}^{\infty} \!\!dy\, G(x,t_i|y,t_{i-1}^+)\rho(y,t^+_{i-1})\Theta(x\!<\!x_c^{(i)}). 
\end{equation}
In the case of forward reshaping, the argument of the $\Theta$ function here and in Eq.~\eqref{EqConsecutivePartialConvolutions} changes to $x\!>\!x_c^{i}$. Note that in Eq.~\eqref{EqConsecutivePartialConvolutions} the harvesting thresholds $x_c^{(i)}$ can be  time independent or set at a fixed percentile of the probability distribution, as we discuss below. Importantly, due to the $\Theta$ function, these consecutive integrations are not associative; namely, unlike the unharvested case, this time $\rho(x,t)\neq \int_0^{\infty}G(x,t|y,0)\rho_0(y)dy$. 

\section{The effects of consecutive harvests}

Consecutive harvests, leading to stochastic reshaping, can be imposed at different time intervals. For simplicity we focus on consecutive time intervals of a fixed duration $\tau$. 
{%
In many cases,  
computing the shape of the probability density $\rho(x,t)$ analytically after a series of consecutive reshapings may be difficult, due to the non-associativity of Eq.~\eqref{EqConsecutivePartialConvolutions}, but it can be solved numerically. 
Notably, in the following we assume that the dynamics is governed by either normal or anomalous diffusion (slower than ballistic) and are bounded by zero from below. The latter is motivated by biological systems in which the $x$-axis represent population size or body size of individuals~\cite{ovaskainen2010stochastic,assaf2017wkb}, which cannot be negative. In order to avoid pathologies, we also assume that the process is not subject to noise with divergent moments~\cite{klafter2011first}.  
Under these conditions  we find two distinct effects, see Fig.~\ref{fig2}: 
\begin{equation}
\label{eq:harvesting_schematic_final}
\begin{tikzpicture}[
    font=\footnotesize,
    every node/.style={align=left, inner sep=1pt, anchor=west},
    arrow/.style={->, line width=1pt},
    x=1cm, y=1cm
]

\node (a) at (0,0.95) {Backward reshaping,\\ below a fixed threshold\\ or below the median};
\node (c) at (0,-0.85) {Forward reshaping,\\ above the median};

\node (e1) at (4.60,1.00) {Effect 1:\\ Convergence to a\\ quasi-steady state};
\node (e2) at (4.60,-0.85) {Effect 2:\\ Convergence to a\\ fixed shape, constant\\ drift toward $\infty$};

\draw[arrow] (3.410,1.00) -- (4.25,1.00);
\draw[arrow] (3.410,-0.85) -- (4.25,-0.85);

\draw[dashed, gray] (-0.1,0.0) -- (5.60,0.0);

\end{tikzpicture}
\end{equation}
In contrast to the effects described in Eq.~\eqref{eq:harvesting_schematic_final}, reshaping forward at a fixed threshold becomes insignificant after several repetitions, since the harvested portion of the  probability density becomes vanishingly small. 
Below, we explain the effects described in Eq.~\eqref{eq:harvesting_schematic_final} in detail. 
 
\subsection{Backward Reshaping}

\textbf{Reshaping backwards at a fixed location.}  
We now provide a heuristic argument for the convergence of the backward-reshaped density to a quasi-static shape, when the harvesting threshold is set at a fixed location, $x_c$, independent of the dynamics (see Appendix~\ref{appendix:fxd_rshp}).    

Consecutive reshapings generate a sequence of $n$ probability distributions; $\rho(x,n\tau)$, in the immediate instant before the $n$th reshaping event ($n = 1, 2,\dots$). Between the reshaping events the system evolves with its original stochastic dynamics. 
\begin{figure}
    \centering
    \hspace{-0.5cm}\includegraphics[width=0.48\textwidth]{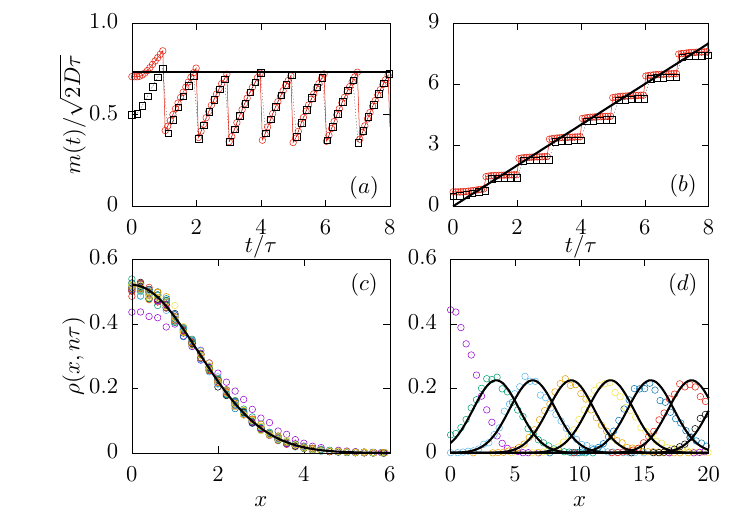}
    \vspace{-0.5cm}\caption{\footnotesize{\textit{Consecutive backward and forward reshaping of Brownian motion.} Normalized median $m(t)/\sqrt{2D\tau}$ versus normalized time $t/\tau$ for
    a Brownian particle starting at $x(0) = 1$ and subject to a reflecting wall at $x = 0$, for $D = 1$. Sampling time:
    $\tau = 1$ (circles) and $\tau = 2$ (squares), for backward-reshaping (a) and forward-reshaping (b),
     with respect to  the median.
    Black lines in (a,b) indicate the analytical predictions, see text below Eq.~\eqref{SteadyApprox} and Eq.~\eqref{EqBMFwReshaping}, respectively.  The position distribution $\rho(x,n\tau)$, right before reshaping events at times $n\tau$ with $n=1,\dots 6$, 
    is shown for the two cases in (c) and (d). In (c), the black line represents 
    Eq.~(\ref{SteadyState}), in perfect agreement with numerical estimates. 
    In (d) the distributions
    are plotted at intervals of $2\tau$ for visible clarity, for $\tau = 1$. Black lines in (d) show the theoretical prediction~\eqref{EqBMFwReshaping}.  }}
    \label{fig2}
\end{figure}  
If a sufficiently large number $N$ of
reshaping events have taken place, and the time interval $\tau$ is finite, then for $n\geq N$, the
elements of the above sequence are monotonically decreasing functions of $x$. Furthermore, as the domains
of all the distributions of the sequence is $[0,x_c]$, the distribution $\rho(x,(n+1)\tau)$ cannot be either
fully above or fully below $\rho_n(x,n\tau)$     for $x\in [0,x_c]$, due to normalization (Appendix~\ref{appendix:fxd_rshp}). This implies
that any pair of consecutive elements of the sequence intersect with each other, with only admissible
limits of this being either the delta peak at $x = 0$ or a uniform distribution over the domain $[0,x_c]$.
As neither of these two limits are admissible following the dynamics of the process, the only possible
scenario is that the successive elements of the sequence that intersect at every point of the domain $[0,x_c]$ are identical.
We find that this sequence of probability density functions converges to a stroboscopic fixed point in time, which can be viewed as a quasi-steady state if the system is observed at time intervals coinciding with the  reshaping events---the harvesting clock ticks.  The exact form of the probability density in this quasi-static state is different for different dynamics, however the existence of this state is universal. Notably, this mechanism is reminiscent of the Yaglom limiting distribution, obtained for random Markov processes conditioned on survival despite an absorbing state~\cite{Sylvie2012Quasi}.
%

\textbf{Reshaping backwards at the median.} Let us now consider 
the scenario where the threshold location of the reshaping, $x_c$, is defined
at a fixed percentile of the distribution $\rho(x,t)$. For simplicity, let $x_c = m(t)$, where $m(t)$ is its median. When the distribution is reshaped backwards  with respect to  the median,
its value changes to the first quartile of the earlier distribution. For a sufficiently large number of reshaping events $N$, the  initial conditions
become irrelevant. As a result, the sequence of medians $\{m(n\tau)\}_{n\geq N}$ approaches a constant
value, see Appendix~\ref{appendix:AppenFoldingBackwards} for details.
This implies that as in the previously discussed case of reshaping backwards with respect
to a fixed location, here too, the sequence of distributions under backward reshaping with respect to  the median
converges to a fixed shape which is monotonically
decreasing with $x$. Yet, reshaping at a fixed location is different from reshaping at the median as in the former the harvesting threshold is externally
imposed, while in the latter the median is state dependent and adaptive.}

\textbf{Finding the quasi-steady state of backward-reshaped Brownian motion.} 
We consider backward reshaping of Brownian motion, with the  threshold  set at the median, and look for the quasi-steady state in the form of a Gaussian truncated at $x\!=\!0$ and renormalized. Denoting the quasi-steady state $\rho_*(x)=\lim_{n\to\infty} \rho(x,n\tau)$ where $n$ is the reshaping index, we assume
\begin{equation}\label{SteadyState}
    \rho_*(x)=[2/(\pi\sigma^2)]^{1/2}\, e^{-x^2/(2\sigma^2)},
\end{equation}
and numerically show that convoluting $\rho_*(x)$ for time $\tau$ remains a half-Gaussian with variance $\sigma^2$. To do so, we use the Green's function with a reflecting boundary at $x\!=\!0$: $
G(x,y,\tau) \!=\!
(4\pi D\tau)^{-1/2}
(e^{-\frac{(x-y)^2}{4D\tau}}\!
+\!
e^{-\frac{(x+y)^2}{4D\tau}})$.
For $\rho_*(x)$ to be a steady state, it must be equal to its convoluted function after being cut off at the median:
\begin{equation}
\rho_*(x) = 2 \int_0^{m^*} G(x,y,\tau)\, \rho_*(y)\, dy.
\end{equation}
Plugging the Green's function, rescaling $\xi=x/\sigma$, $\Upsilon=y/\sigma$ and $m^*=a\sigma$, and writing $\sigma=A\sqrt{D\tau}$, we find
\begin{eqnarray}
&&\hspace{-9mm}\rho_*(\xi)=
\frac{\sqrt{2}}{\sqrt{\pi^2 D\tau}}\int_0^a
\!\!\left(
e^{-\frac{A^2 (\xi\!-\!\Upsilon)^2}{4}}
\!+\!
e^{-\frac{A^2 (\xi\!+\!\Upsilon)^2}{4}}
\right)
e^{-\frac{\Upsilon^2}{2}}
\, d\Upsilon\nonumber\\
&&\hspace{-9mm}=
\frac{\sqrt{2}A}{\sqrt{\pi(2\!+\!A^2)}}
e^{-\frac{\xi^2}{2+\frac{4}{A^2}}}
\left(
\operatorname{erf}(B\!-\!C\xi)
\!+\!
\operatorname{erf}(B\!+\!C\xi)
\right)\!,
\label{SteadyApprox}
\end{eqnarray}
where $\operatorname{erf}$ is the error function,
$B = (a/2)\sqrt{2+A^2}$, and $C = (A^2/2)/\sqrt{2+A^2}$. 
Equation~(\ref{SteadyApprox}) is 
indistinguishable from the initial half Gaussian ansatz.
To find $a$ we demand that $\int_0^{m^*}
\sqrt{2/(\pi \sigma^2)}e^{-\frac{y^2}{2\sigma^2}}dy=1/2$, 
which yields $a=0.67449$---the Gaussian's 75th percentile.
To find $A$---the width of the half-Gaussian---we demand that 
\begin{equation}
\!\!\frac{1}{2}
\!=\!
\int_0^a\!\!\!
\frac{\sqrt{2}A}{\sqrt{\pi(2\!+\!A^2)}}
e^{-\frac{\xi^2}{2+\frac{4}{A^2}}}
\!\!\left(
\operatorname{Erf}(B\!-\!C\xi)
\!+\!
\operatorname{Erf}(B\!+\!C\xi)
\right)\!\!d\xi.\nonumber
\end{equation}
Solving it numerically, yields $A = 1.53229$.
That is, the half Gaussian steady-state solution of the backward reshaping has mean zero and width $\sigma\simeq 1.5323\sqrt{D\tau}$.

Figures~\ref{fig2}(a) and~\ref{fig2}(c) show the excellent agreement between Brownian motion simulation results and the analytical predictions. In panel (a), the median, obtained for two different parameter sets grows diffusively between the reshaping events, and the different simulation results match when normalizing by $\sqrt{2D\tau}$. After $n\gg1$ reshapes $m(n\tau)/\sqrt{2D\tau}$ approaches the black line at $m^*\!/\!\sqrt{2}\!\approx\! 0.731$ (see calculation above).  In (c), the initial spatial density is shown in purple, and plotted in different colors after six reshaping+diffusion cycles, converging to the black line corresponding to the quasi-steady state [Eq.~\eqref{SteadyApprox}].

\vspace{-0.2cm}\subsection{Forward Reshaping}
\vspace{-0.2cm}
\textbf{Forward Reshaping at the median.} 
In forward reshaping with respect to the median, the median changes to the third quartile; $q_3(t)$: $m((n\tau)^+) 
= q_3((n-1)\tau)~\forall~n\geq 1$. The time $(n\tau)^+$ is the immediate instant after the $n$th reshaping event. Analogous to the case of reshaping backwards, after a
sufficiently large number $N$ of reshaping events,  the initial conditions become
irrelevant. As explained in detail in Appendix~\ref{appendix:AppenFoldingForward},
reshaping forward with respect to the median removes random walkers which are found closer to the wall. As a result, their spatial density becomes broader and its percentiles shift farther away from $x=0$ than the density following the previous reshaping event. Without reshaping, the spatial density naturally expands away from the origin,  
and as forward reshaping events follow the same direction, they further push
the distribution rightward, leading to a feedback loop that  
causes it to converge to a fixed shape and drift towards $x\rightarrow\infty$ at a fixed pace. Note that, forward reshaping reminds of the cutoff-front mechanism of Brunet and Derrida \cite{brunet1997shift}, and fitness waves of evolutionary dynamics in branching stochastic processes~\cite{tsimring1996rna,desai2007beneficial,hallatschek2011noisy}. In this sense, the procedure reads as branching selection stripped of its branching: diffusion supplies the spread, and truncation converts it into an effective drift pushing the front forward. 

\textbf{Finding the probability density of forward-reshaped Brownian motion.} While the analytical formula for a general process is infeasible, in Appendix~\ref{AppenApproxBMFwReshaping} we derive an asymptotic solution in the case of Brownian motion. 
Here, after $n$ reshaping events, we find: 
\begin{eqnarray}
    \Delta m&\!\equiv\! m((n+1)\tau)\!-\!m({n\tau})\!=\!\sqrt{2/\pi}\sigma(n\tau)\xrightarrow[n\to\infty]{}\!\sqrt{2D\tau},\nonumber\\ &\hspace{-1.4cm}\sigma((n+1)\tau)^2=\pi D\tau \left[1-\left(1-2/\pi\right)^n\right]\xrightarrow[n\to\infty]{}\pi D\tau. 
    \label{EqBMFwReshaping}
\end{eqnarray} 
As shown in the appendix, this result is valid for a long diffusion step, i.e., $D\tau\gtrsim 1$, which indicates that, regardless of the state of the density prior to reshaping, after the diffusion evolution, the distribution's bulk is well described by a Gaussian. In Fig.~\ref{fig2}(b,d), we demonstrate how the spatial density converges to a fixed shape, while its median shifts to the right at constant rate. 
 The mismatch between the simulation and analytical results arises because the probability distribution's tails are in fact not symmetric, see Appendix~\ref{AppenApproxBMFwReshaping}. 

\section{Anomalous diffusion} 
\label{SecPhenomenon}
\begin{figure}
    \centering
    \hspace{-0.55cm}\includegraphics[width=0.5\textwidth]{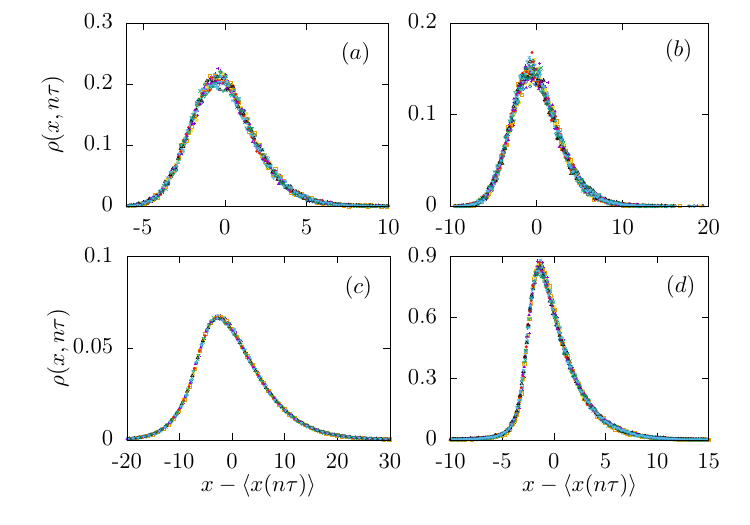}
    \vspace{-0.6cm}\caption{
\footnotesize{\textit{Generality of forward-reshaping: from sub- to superdiffusion.} Position distribution $\rho(x,n\tau)$ shifted by the mean $\langle x(n\tau)\rangle$
    prior to reshaping events for the case when the distribution
    is reshaped forward  with respect to  the median for different random walks starting at $x(0) = 1$ and subject to a reflecting wall at $x = 0$. The independent random walks are: 
(a) Brownian motion, (b) superdiffusive run-and-tumble dynamics, (c) subdiffusive continuous-time random walks, and (d) superdiffusive L\'evy walks. 
(see Appendix~\ref{appendix:reshape} and \ref{appendix:brwn} for details). 
    }}
    \label{fig3}
\end{figure}

\vspace{-0.2cm}
To extend the generality of our results to beyond Brownian motion, we numerically study three paradigmatic models of anomalous diffusion, including both subdiffusion and superdiffusion. 
 Figure.~\ref{fig3}  shows the results of consecutive forward-reshaping applied to
(a) Brownian particles with diffusion coefficient
$D = 1$ and time-interval for reshaping $\tau = 1$ as the baseline for comparison, and (b) run-and-tumble particles~\cite{DharRun-and-tumble}, with particle velocity $v_0 = 1.5$, 
    rate of tumbling $r = 0.8$, diffusion coefficient $D = 0.5$, and time-interval between consecutive reshapings; $\tau = 4$.
    Panel (c) shows subdiffusive continuous-time random walk (CTRW)~\cite{ScherMontroll} on a lattice, with a waiting time distribution
    $\psi(T) \sim 1/T^{1+\alpha}$ with $\alpha = 0.2$ and time-interval $\tau = 10^3$,
    and (d) presents superdiffusive L\'evy walks~\cite{zaburdaev2015levy}, with velocity $v_0 = 1$ and flight time distribution $\psi(T) \sim
    1/T^{1+\alpha}$ with $\alpha = 1.5$ and $\tau = 10$.
  In all the cases, the independent random walkers start from a $\delta$-distribution centered at $x(0) = 1$. The image shows  that after several reshapes the probability density in all these processes converges to a fixed shape (after rescaling time to remove the linear drift), though the exact shape depends on the processes involved.

Note that, for the unharvested dynamics, for both Brownian and  run-and-tumble particles the median grows as $m(t) \sim t^{1/2}$; yet, for subdiffusive CTRW, $m(t) \sim t^{\alpha/2}$ with $0<\alpha<1$, and for superdiffusive
L\'{e}vy walk  $m(t) \sim t^{(3-\alpha)/2}$ with $1<\alpha<2$. For these processes, both the median and mean of the unharvested system grow more slowly than ballistic. 
For the forward-harvested systems, when observed at successive ticks of the harvesting clock and shifted by the instantaneous mean, the spatial densities collapse onto a single curve in all cases, confirming that the density shape is fixed while the mean advances linearly in time.
{The generality of the convergence to a quasi-steady state due to repeated backward reshaping is demonstrated in Appendix~\ref{appendix:AppenFoldingBackwards}.} 

\section{Harvesting in a predator-prey model}
\vspace{-0.2cm}
Consider a predator-prey toy model, where the predators have a distribution of physical size, $x$; $\rho(x,t)$. The larger the predator, the better it is at hunting. 
For simplicity, we assume that the size of the predators varies randomly, e.g., due to birth, death and migration, so its dynamics is affected by white noise. The prey abundance at time $t$ is $y(t)$. This system is described by the Langevin equations: 
\begin{align}
& \dot{x}(t) = \sigma \Gamma(t), \quad \text{with a reflecting boundary at } x=0, \nonumber \\
& \dot{y}(t) = b y(t) - \langle x(t)\rangle y(t),
\label{EqPredatorPreyLangevins}
\end{align}
where $\Gamma(t)$ is zero-mean, unit-variance, Gaussian white noise and $\sigma>0$. 
The term $b y(t)$ ($b>0$) represents prey reproduction, while $\langle x(t)\rangle y(t)$ accounts for predation, proportional to the mean predator size. 

Starting from a narrow initial probability density function of predator sizes, this density expands over time as a half-Gaussian on $x>0$. 
As a result, the mean predator size grows as $\langle x(t)\rangle=\sqrt{\sigma t}$, until  ultimately predation surpasses prey reproduction. 
To counteract this effect, predators can be externally culled at regular time intervals by removing all individuals larger than the median size~\footnote{This simplified model  implicitly assumes that the predator population size quickly recovers between harvests, due to e.g., predation of other prey besides $y$.}. Note that, the normalization of the probability density after reshaping does not affect the dynamics, since the latter only depends on the mean predator body size rather than the predator abundance.

\begin{figure}
    \centering
    \hspace{-0.45cm}\includegraphics[width=0.47\textwidth]{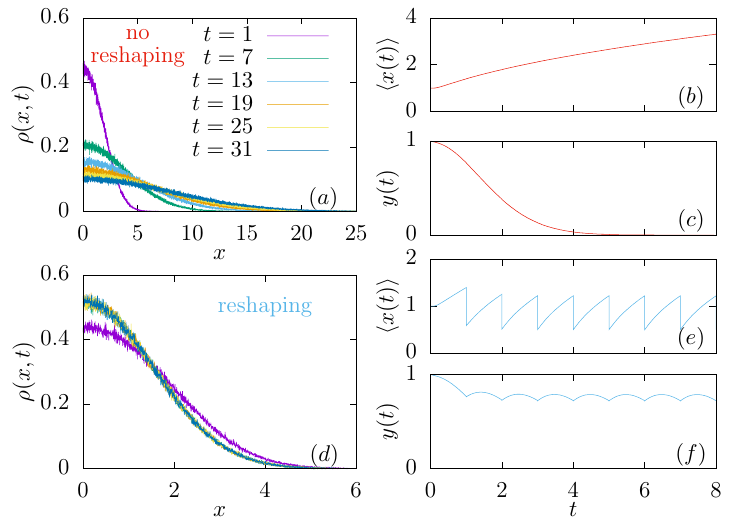}
    \vspace{-0.5cm}\caption{\footnotesize{\textit{Harvesting in a predator-prey system.} (a,b) Temporal evolution of an unharvested population of predators; $\rho(x,t)$ and its mean $\lr x(t) \rl$,
    and (c) its effect of the population of the prey $y(t)$. (d-e) Distribution of predator sizes under consecutive reshapings 
    at time intervals of duration $\tau$, and its affect on the corresponding mean predator size. (f)~The effect of the predator density reshaping, on the prey population.
    Initially, $x(0) = 1$ and $y(0) = 1$, and the evolution is according to Eq.~\eqref{EqPredatorPreyLangevins}. The simulation parameters
    are: $\tau = 1,~\sigma = 1,~b = 0.911$, see Appendix~\ref{appendix:reshape} and~\ref{appendix:brwn}.}}
\label{FigHarvestingPredatorPrey}
\end{figure}

Figure~\ref{FigHarvestingPredatorPrey} shows the effect of consecutive harvesting events on the predator and prey populations. 
Without harvesting, at some point in time the prey become out-competed by the predators, and its population goes extinct as indicated by the numerical simulations of Eq.~\eqref{EqPredatorPreyLangevins} (Fig.~\ref{FigHarvestingPredatorPrey}(a-c)). However, when the predator population undergoes repeated backward reshapings, its distribution $\rho(x,t)$ and in particular, its mean $\lr x(t) \rl
= \int^\infty_0 dx~x \rho(x,t)$ are modified. This allows the prey population to recover after each harvest and ultimately leads to a quasi-steady state of coexistence of both species.
Notably,  repeated density reshapings drive the predator size distribution toward a fixed form when observed at the instants right before each harvest.

\section{Discussion} 

\textit{Stochastic reshapings} are a macroscopic intervention in which the evolution of the system is altered through its distribution, without affecting the underlying dynamics. 
One well known method to steer stochastic dynamics is through stochastic resetting \cite{evans2011diffusion,evans2020stochastic,gupta2022stochastic}. As opposed to resetting, reshaping does not change the particle positions. Instead, it can be achieved by removal of elements from the system. The parts of the system which are not harvested out continue to evolve unaltered. Our focus on systems bounded from below is motivated by biological systems where the fluctuating quantity can be the animal size or population abundance~\cite{ovaskainen2010stochastic,assaf2017wkb}. This lower bound gives rise to two very distinct reshaping protocols, unlike systems that can expand symmetrically to $\pm\infty$.

Repeated backward reshapings of continuous-time, 
Markovian stochastic dynamics 
drives the system toward a quasi-steady state: removing walkers beyond the median preserves the ``losing'' portion of the population near the origin, stalling the overall spatial expansion without collapsing the distribution. 
In contrast, forward reshaping leads the density to a fixed shape while inducing an effective drift toward $x\to\infty$. 
By repeatedly exploiting the most extreme stochastic excursions, this protocol drives the population forward faster than in an unharvested Brownian packet, whose mean scales as $\langle x(t)\rangle \propto\sqrt{t}$. 
Under harvesting, the scaling of the mean particle position is replaced by a linear drift, $\langle x(t)\rangle\propto t$.

\begin{figure}
    \centering
    \hspace{-0.5cm}\includegraphics[width=0.45\textwidth]{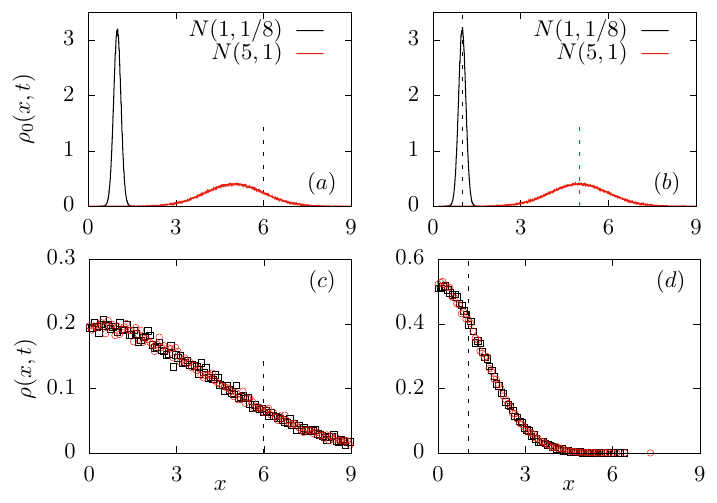}
    \vspace{-0.5cm}\caption{\footnotesize{\textit{The quasi-steady state in backward reshaping.} Panels (a,b): The initial spatial density of Brownian particles, prior to backward reshaping 
      with respect to (a) a fixed location at $a = 6$ and (b) the median. The initial Gaussian density in black has unit mean and variance $1/8$, and the mean and variance for the red density are $5$ and $1$. The reshaping thresholds in each panel are marked with dashed lines. In (a) there is a single reshaping threshold, while in (b) the threshold differs between the two initial densities. In panels (c,d) shown is the distribution of the Brownian particle just before the fifth reshaping event, for reshaping with respect to a fixed threshold (c) (similar to that in panel (a)) and median (d). Note that, in (d), the two initially different thresholds converge to a single value upon reaching a quasi-steady state after multiple reshapings.  
      The total simulation time is $t=5$, $\tau = 1$, the number of particles is $10^6+1$, and $D = 1$ (Appendix~\ref{appendix:reshape} and \ref{appendix:brwn}).}}
    \label{intl_cond}
\end{figure}

In the processes we have examined, our numerical investigation suggests that the system's state after backward reshaping depends on the duration of the time interval between the reshapings and the threshold, but not on the initial state of the system.  
In Fig.~\ref{intl_cond} we check the effect of choosing a fixed reshaping threshold versus reshaping at the median, and the effect of the initial conditions on the final density for Brownian motion.  
In (a) and (b) we present the initial conditions: one narrow and one broad Gaussian distribution. Dashed lines mark the
location of the reshaping threshold. In (a) the threshold is fixed: $x_c = 6$, whereas in (b), the reshaping threshold is set at the median of the initial density. Panels (c,d) demonstrate that the quasi-steady state arising due to reshapes is independent of the initial distribution, but not the reshaping protocol. The number of reshape events required to reach the quasi-steady state does depend on the initial condition: both in (c) and (d),  to converge to an approximately fixed shape required two reshapings for the narrow initial state and four for the broad one. The density shapes are in agreement with Eq.~\eqref{SteadyApprox}.

We have also numerically confirmed that  reshaping the probability density backwards with respect to any 
percentile also leads to a convergence to a quasi-steady state, albeit different from the shape resulting from median reshaping. Similarly, forward-reshaping with respect to any percentile ultimately fixes the shape of the density and introduces a linear drift, as  for median reshaping. 

\acknowledgments
We thank E. Frey for useful comments. J.R.G. acknowledges support from the U.S. Department of Energy, Office of Science, Office of Basic Energy Sciences, Funding for Accelerated, Inclusive, Research (FAIR) under Award No. DE-SC-0024305.  
\endacknowledgments

\bibliography{bibliography}

@article{ScherMontroll,
  title = {Anomalous transit-time dispersion in amorphous solids},
  author = {Scher, Harvey and Montroll, Elliott W.},
  journal = {Physical Review  B},
  volume = {12},
  issue = {6},
  pages = {2455--2477},
  numpages = {0},
  year = {1975},
  month = {Sep},
  publisher = {American Physical Society},
  doi = {10.1103/PhysRevB.12.2455},
  url = {https://link.aps.org/doi/10.1103/PhysRevB.12.2455}
}

@article{DharRun-and-tumble,
  title = {Run-and-tumble particle in one-dimensional confining potentials: Steady-state, relaxation, and first-passage properties},
  author = {Dhar, Abhishek and Kundu, Anupam and Majumdar, Satya N. and Sabhapandit, Sanjib and Schehr, Gr\'egory},
  journal = {Physical Review  E},
  volume = {99},
  issue = {3},
  pages = {032132},
  numpages = {14},
  year = {2019},
  month = {Mar},
  publisher = {American Physical Society},
  doi = {10.1103/PhysRevE.99.032132},
  url = {https://link.aps.org/doi/10.1103/PhysRevE.99.032132}
}

@article{thambithurai2024environmental,
  title={Environmental forcing alters fisheries selection},
  author={Thambithurai, Davide and Kuparinen, Anna},
  journal={Trends in Ecology \& Evolution},
  volume={39},
  number={2},
  pages={131--140},
  year={2024},
  publisher={Elsevier}, 
doi = {10.1016/j.tree.2023.08.015}
}

@article{charmantier2024does,
  title={How does urbanization affect natural selection?},
  author={Charmantier, Anne and Burkhard, Tracy and Gervais, Laura and Perrier, Charles and Schulte-Hostedde, Albrecht I and Thompson, Megan J},
  journal={Functional Ecology},
  volume={38},
  number={12},
  pages={2522--2536},
  year={2024},
  publisher={Wiley Online Library}, 
doi= {10.1111/1365-2435.14667}
}

@article{darimont2009human,
  title={Human predators outpace other agents of trait change in the wild},
  author={Darimont, Chris T and Carlson, Stephanie M and Kinnison, Michael T and Paquet, Paul C and Reimchen, Thomas E and Wilmers, Christopher C},
  journal={Proceedings of the National Academy of Sciences},
  volume={106},
  number={3},
  pages={952--954},
  year={2009},
  publisher={National Academy of Sciences}, 
doi={10.1073/pnas.0809235106} 
}

@article{evans2011diffusion,
  title={Diffusion with stochastic resetting},
  author={Evans, Martin R and Majumdar, Satya N},
  journal={Physical review letters},
  volume={106},
  number={16},
  pages={160601},
  year={2011},
  publisher={APS}
}

@article{brunet1997shift,
  title={Shift in the velocity of a front due to a cutoff},
  author={Brunet, Eric and Derrida, Bernard},
  journal={Physical Review E},
  volume={56},
  number={3},
  pages={2597},
  year={1997},
  publisher={APS}
}

@article{hallatschek2011noisy,
  title={The noisy edge of traveling waves},
  author={Hallatschek, Oskar},
  journal={Proceedings of the National Academy of Sciences},
  volume={108},
  number={5},
  pages={1783--1787},
  year={2011},
  publisher={National Academy of Sciences}
}

@article{desai2007beneficial,
  title={Beneficial mutation--selection balance and the effect of linkage on positive selection},
  author={Desai, Michael M and Fisher, Daniel S},
  journal={Genetics},
  volume={176},
  number={3},
  pages={1759--1798},
  year={2007},
  publisher={Oxford University Press}
}

@article{tsimring1996rna,
  title={RNA virus evolution via a fitness-space model},
  author={Tsimring, Lev S and Levine, Herbert and Kessler, David A},
  journal={Physical review letters},
  volume={76},
  number={23},
  pages={4440},
  year={1996},
  publisher={APS}
}

@article{burdzy2000fleming,
  title={A Fleming--Viot Particle Representation of the Dirichlet Laplacian},
  author={Burdzy, Krzysztof and Ho{\l}yst, Robert and March, Peter},
  journal={Communications in Mathematical Physics},
  volume={214},
  number={3},
  pages={679--703},
  year={2000},
  publisher={Springer}
}

@article{Sylvie2012Quasi,
author = {Sylvie M{\'e}l{\'e}ard and Denis Villemonais},
title = {{Quasi-stationary distributions and population processes}},
volume = {9},
journal = {Probability Surveys},
number = {none},
publisher = {Institute of Mathematical Statistics and Bernoulli Society},
pages = {340 -- 410},
year = {2012},
doi = {10.1214/11-PS191},
URL = {https://doi.org/10.1214/11-PS191}
}

@article{metzler2000random,
  title={The random walk's guide to anomalous diffusion: a fractional dynamics approach},
  author={Metzler, Ralf and Klafter, Joseph},
  journal={Physics reports},
  volume={339},
  number={1},
  pages={1--77},
  year={2000},
  publisher={Elsevier}, 
doi={10.1016/S0370-1573(00)00070-3
Get}
}

@article{ovaskainen2010stochastic,
  title={Stochastic models of population extinction},
  author={Ovaskainen, Otso and Meerson, Baruch},
  journal={Trends in ecology \& evolution},
  volume={25},
  number={11},
  pages={643--652},
  year={2010},
  publisher={Elsevier}
}

@article{assaf2017wkb,
  title={{WKB} theory of large deviations in stochastic populations},
  author={Assaf, Michael and Meerson, Baruch},
  journal={Journal of Physics A: Mathematical and Theoretical},
  volume={50},
  number={26},
  pages={263001},
  year={2017},
  publisher={IOP Publishing}
}

@book{toral2014stochastic,
  title={Stochastic Numerical Methods: An Introduction for Students and Scientists},
  author={Toral, R. and Colet, P.},
  year={2014},
  publisher={John Wiley \& Sons}, 
doi={10.1002/9783527683147}
}

@book{klafter2011first,
  title={First Steps in Random Walks: From Tools to Applications},
  author={Klafter, J. and Sokolov, I. M.},
  year={2011},
  publisher={OUP Oxford}, 
doi={10.1093/acprof:oso/9780199234868.001.0001}
}

@article{malakar2018steady,
  title={Steady state, relaxation and first-passage properties of a run-and-tumble particle in one-dimension},
  author={Malakar, K. and Jemseena, V. and Kundu, A. and Kumar, K. V. and Sabhapandit, S. and Majumdar, S. N. and Redner, S. and Dhar, A.},
  journal={Journal of Statistical Mechanics: Theory and Experiment},
  volume={2018},
  number={4},
  pages={043215},
  year={2018},
  publisher={IOP Publishing}, 
doi={10.1088/1742-5468/aab84f}
}

@article{zaburdaev2015levy,
  title={L{\'e}vy walks},
  author={Zaburdaev, V. and Denisov, S. and Klafter, J.},
  journal={Reviews of Modern Physics},
  volume={87},
  number={2},
  pages={483--530},
  year={2015},
  publisher={APS}, 
doi={10.1103/RevModPhys.87.483
}}

@article{evans2020stochastic,
  title={Stochastic resetting and applications},
  author={Evans, M. R. and Majumdar, S. N. and Schehr, G.},
  journal={Journal of Physics A: Mathematical and Theoretical},
  volume={53},
  number={19},
  pages={193001},
  year={2020},
  publisher={IOP Publishing}, 
doi={10.1088/1751-8121/ab7cfe}
}

@article{gupta2022stochastic,
  title={Stochastic resetting: A (very) brief review},
  author={Gupta, S. and Jayannavar, A. M.},
  journal={Frontiers in Physics},
  volume={10},
  pages={789097},
  year={2022},
  publisher={Frontiers Media SA}, 
doi={10.3389/fphy.2022.789097}
}

\appendix
\setcounter{secnumdepth}{1}
\renewcommand{\thesection}{\Alph{section}}
\renewcommand{\theHsection}{\thesection} 

\section{Algorithm for simulating stochastic processes  under reshaping about the median}
\label{appendix:reshape}

The probability distribution of a set of stochastic processes is evolved in time, either by iterating over the dynamical equations of motion of individual system elements, or by iterating over the the density dynamics using its evolution kernel. We define stochastic reshaping as follows: 
at regular time intervals, the density is clipped instantaneously from above, or below the reshaping threshold---be it a fixed value of a percentile of its current state. In an agent-based simulation, this is achieved by removing all the particles found above/below the threshold, depending on the desired protocol. Upon simulating the macroscopic dynamics, the probability density is clipped and renormalized as in Eq.~\eqref{EqConsecutivePartialConvolutions}. Following a reshaping event, the dynamics proceeds as before. 

We now present the specific agent-based algorithm, used to numerically study random walks under reshapes with respect to the median. Here, we focus on random walks with a subject to a reflecting
    wall at $x = 0$. This algorithm extends straightforwardly to reshaping at any percentile or a fixed harvesting threshold, for any stochastic dynamics. Note that for numerical convenience, in simulations, we fix the number of particles by replacing the harvested ones with new particles, whose random locations are selected from the renormalized density. For sufficiently large populations, this is identical to re-weighting the remaining population after each harvest in Eq.~\eqref{EqReshaping}, where particles are removed  without replacement. 

Consider a collection of $N$ independent random walks all starting at the same initial location
    $x = a (>0)$. 
\begin{enumerate}
    \item The positions of the random walks evolve following its equation of motion for a time interval $\tau$, subject to a reflecting
    wall at $x = 0$.
    \item Sample the positions of the random waler at $\tau$.
    \item Find the median walker positions, $m(\tau)$. 
    Then, the positions
    are reshaped with respect to the median, such that the new positions $\{y_1(\tau),\dots,y_N(\tau)\}$ are:
    \begin{align}
        y_i(\tau) = \begin{cases}
        x_i(\tau),~\text{for}~i=1,\dots,N/2+1,\\
        x_{N-i+1}(\tau),~\text{for}~i=N/2+2,\dots,N,
        \end{cases}
    \end{align}
    if the positions above the median are reshaped. In the alternative case, when the positions of the
    random walks below the median are reshaped then the above equation modifies to
    \begin{align}
        y_i(\tau) = \begin{cases}
        x_{N-i+1}(\tau),~\text{for}~i=1,\dots,N/2+1,\\
        x_i(\tau),~\text{for}~i=N/2+2,\dots,N.
        \end{cases}
    \end{align}
    It is to be noted here that $N$ is chosen to be odd.
    \item The positions $\{y_1(\tau),\dots,y_N(\tau)\}$ are now chosen to be the initial locations of the
    random walks with time initialized to $t = \tau$.
    \item Steps $2-5$ are repeated for $k$ iterations such that $k\tau \leq T$, with $T$ being the walk's duration.
\end{enumerate}
Distribution of the positions $\{x_1(n\tau),\dots,x_N(n\tau)\}$ are measured at every iteration.


\section{Stochastic simulation details}
\label{appendix:brwn}
\vspace{-0.3cm}\textbf{Simulation of Brownian particles.} 
The dynamics is simulated using the Euler-Maruyama method \cite{toral2014stochastic}:
\begin{align}
\label{brwn_dyn}
x(t+dt) = x(t) + \sqrt{2D dt}\eta(t),
\end{align}
where $\eta(t)$ is Gaussian white noise, with zero mean and unit variance. As the Brownian particles evolve independently, we solve a system of
$N$ such stochastic differential equations to study the distribution $\rho(x,t)$, and its subsequent
properties under reshapes. To apply this method to simulate the predator-prey system, we solve
the coupled set of stochastic differential equations in Eq.~(\ref{EqPredatorPreyLangevins}).

\textbf{Run and Tumble random walk.}
Assuming that the initial velocity of a run-and-tumble particle is chosen from the set $\{-v_0,v_0\}$ uniformly, the evolution of its position is described by a set of equations similar to \cite{malakar2018steady}, with instantaneous orientation changes. In practice, ``instantaneous'' means that we consider the duration of the re-orientations, during which the particle's position does not change, as $dt\lll1$. The velocity flips occur with probability $r dt$, where $r$ is the tumbling rate. The equation of motion is thus:
\begin{align}
\label{rtp_dyn}
x(t+dt) = \begin{cases}
x(t) + \sqrt{2Ddt}\eta(t) + v_0 dt,~\text{with probability}\\
~~~~~~~~~~~~~~~~~~~~~~~~~~~~~~~~~~~1-r dt,\\
x(t),~\text{with probability $r dt$}.
\end{cases}
\end{align}
Here, $D$ is its diffusion coefficient. 
As in the previous case of Brownian particles, we solve Eq.~(\ref{rtp_dyn}) for $N$ independent run-and-tumblers to estimate $\rho(x,t)$.

\textbf{Continuous time random walk.} 
The dynamics of CTRW is described by these iterative equations
\cite{metzler2000random}:
\begin{equation} 
\label{ctrw_dyn}
x_{n+1} = x_n + \Delta x,\quad\quad
t_{n+1} = t_n + T,
\end{equation}
where $\Delta x \in \{-1,1\}$ with equal probability and the waiting time $T$ follows from the distribution
$\psi(T)$. CTRW exhibits subdiffusive transport for power-law decaying distribution of waiting times,
that is, $\psi(T) \sim 1/T^{1+\alpha}$ with $\alpha < 1$ \cite{klafter2011first}.
In Eqs.~(\ref{ctrw_dyn}), $n\geq 0$ denotes the
number of jumps and $\Delta x$ and $T$ are independent and identically distributed random variables with
their respective distributions. For a system of independent CTRWs, we solve $N$ such iterative equations.

\textbf{L\'{e}vy walk.} 
A L\'{e}vy walk is analogous to run-and-tumble, but with $D = 0$, and describes motion of particle traveling at a fixed
velocity for a given interval of time, after which it tumbles to a new direction \cite{zaburdaev2015levy}.
However, to extend Eq.~(\ref{rtp_dyn}) to L\'{e}vy
walks with arbitrary distribution of flight times $T$, we use the following set of equations:
\begin{align}
\label{levy_dyn}
x_{n+1} = x_n + v_0 T,
\end{align}
where $n\geq 0$ is the number of directional changes in velocity $v_0$.
We assume that initially $v_0$ is uniformly chosen from the set $\{-1,1\}$. This
L\'{e}vy walk exhibits superdiffusion for a power-law distribution
of flight times: $\psi(T) \sim 1/T^{1+\alpha}$ with $1<\alpha<2$ \cite{klafter2011first}. Using
the above equation, $N$ independent L\'{e}vy walks can be simulated.

\section{Argument for the steady state existence for backward reshaping at a fixed threshold}
\label{appendix:fxd_rshp}
\vspace{-0.3cm}
After a large number of reshaping events have taken place, the elements of the sequence of probability densities $\{\rho(x,(n\tau)^+)\}_{n\geq N}$ satisfy the following 
properties:
\begin{align} 
\label{pr_prob}
&\hspace{-0.125cm}\int^{x_c}_0 dx~\rho(x,(n\tau)^+) = 1,~\text{normalization},\nonumber\\
&\hspace{-0.125cm}\frac{d}{dx}\rho(x,(n\tau)^+)\Big|_{x=0} = 0,~\text{reflecting wall at the origin},\nonumber\\
&\hspace{-0.125cm}\frac{d}{dx}\rho(x,(n\tau)^+) < 0~\forall~x\in(0,x_c],~\text{due to the wall}. 
\end{align}
Note that while the first two properties in Eq.~(\ref{pr_prob}) hold true for every
$n \geq 1$, the last property is satisfied beyond a sufficient number of reshapes, when the effect of
initial conditions are no longer present. The normalization condition implies that every term in the density sequence must be either always larger than the previous one, for every $x\in(0,x_c]$, or always smaller, depending on the evolution's duration between reshapings, unless all the terms are identical. Namely, the density function after each reshape event is either always larger or always smaller than the function resulting from the previous reshaping, if the sequence has not converged to a single shape. 
However, if each density function after each reshaping event were larger than the density following the previous reshaping, everywhere in the range, this would imply that $\rho(x,t)$ approaches a $\delta-$function shape near the wall. This would be impossible, since the tendency of the system is to expand away from the wall between the reshapings. Alternatively, if every term in the set of density function right after the reshapings were smaller than its predecessor, for every $0<x<x_c$, it would mean that $\rho(x,t)$ becomes flat in that region, instead of monotonically decreasing (Eq.~\eqref{appendix:fxd_rshp}). 
As a result, one must conclude that the density functions in the sequence $\{\rho(x,(n\tau)^+)\}_{n\geq N}$, must converge to a fixed shape. 

\section{Argument for the steady state existence due  to backward reshaping at the median} 
\label{appendix:AppenFoldingBackwards} 
\vspace{-0.3cm}
Define the complementary cumulative distribution function (CCDF); 
$F(x,\tau^+) = \int^\infty_x dy~\rho(x,\tau^+)$. After the first time the distribution is reshaped backward with respect to  the median $m(\tau)$, the CCDF is:
\begin{align}
\label{ccdf_bck}
F(x,\tau^+) = \begin{cases}
\int^{m(\tau)}_x dy~\rho(y,\tau^+),~x\leq m(\tau),\\
0,~\text{otherwise}.
\end{cases}
\end{align}
This process is repeated at times $t = n\tau,~n = 1, 2, \dots$ and after
every reshaping event, the median under reshaping shifts back to the first quartile $m((n\tau)^+) = q_1((n-1)\tau)$.
While the above statement holds true for every $n\geq 1$, let us assume that a sufficient
number $N$ of reshaping events have taken place so that dynamics now is almost entirely independent of 
the initial condition. Under these conditions, for $t \in ((n\tau)^+,
(n+1)\tau)$, $F(m(n\tau),t)\leq F(m(n\tau),n\tau) = 1/2~\forall~n\geq N$.
This is because if prior to a reshaping event,
$\text{Pr}(x \in [0,m(n\tau)]) = 1/2$ in a time-interval $\tau$, then conservation of probability
over the semi-infinite line implies that the renormalized distribution at $t = (n\tau)^+$ loses at most
half of its probability in the region $x \in [0,m(n\tau)]$. Thus,
the sequence of medians post-reshape: $\{m((n\tau)^+)\}^\infty_N$ satisfies the
 inequality:
$m((n\tau)^+) \leq m(((n-1)\tau)^+)$.
If the above inequality is strict, then this would imply that the above sequence of medians will
have a limit zero, as the motion is bounded by the reflecting wall at $x = 0$. As a result, 
the increment $m((n+1)\tau)-m((n\tau)^+)$ would approach zero as a function of $n$ implying that
the random walk does not evolve over the finite time interval $t \in ((n\tau)^+,(n+1)\tau)$.

But, the above would be in direct contradiction with the dynamics of the random walk, which evolves naturally away from the wall between the reshaping events. Therefore,  the sequence of medians $\{m(n\tau)\}_{n\geq N}$ must converge to a fixed value
for $n\geq N$, instead of decreasing.
Since the following logic extends to any percentile of the probability density immediately after reshaping, we conclude that as in the case of backward reshaping at a fixed threshold, the functions in the set $\{\rho(x,n\tau)\}_{n\geq N}$
must converge to a fixed shape.

\begin{figure}
    \centering
    \includegraphics[width=0.43\textwidth]{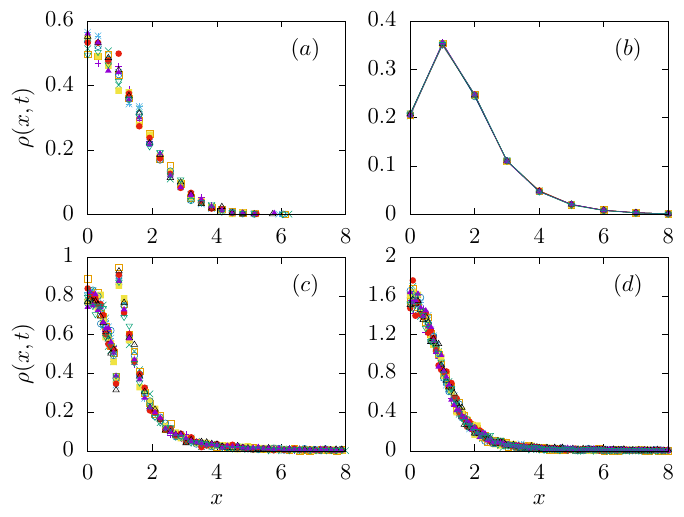}
    \vspace{-0.5cm}\caption{\footnotesize{\textit{Convergence of the probability density of anomalous random walkers to a quasi-steady state.}} Distribution $\rho^r(x,t)$
    for (a) run-and-tumble particles, (b) continuous time random walk, and (c,d) L\'{e}vy walks, in the case of backward reshaping. In (c), parameter
    values are same as that in Fig.~\ref{fig3}. In (d) the parameters similar, but the L\'{e}vy walks start at
    $x(0) = 1.15$.}
    \label{fig_bck}
\end{figure}
To demonstrate the generality of this effect, in Fig.~\ref{fig_bck} we study the properties of backward reshaping at the median for both subdiffusive and superdiffusive random walks, and observe that analogous to Brownian particles (see main text), the position
distribution of random walks exhibiting anomalous diffusion also converges to a nonequilibrium steady
state under backward reshaping with respect to  the median. Note, however,
the peculiar case of L\'{e}vy walks in Figs.~(\ref{fig_bck})(c) and (d), wherein a jump in the position
distribution is observed for L\'{e}vy walks starting at $x(0) = 1$, but missing for those starting
at $x(0) = 1.15$. We attribute the reason for this difference to be to commensurability of $x(0)$ and $v_0/\tau$ (in these two examples $v_0=1$) for
L\'{e}vy walks, though further investigation is needed to fully uncover the details.

\vspace{-0.2cm}

\section{Argument for the existence of a fixed shape in forward reshaping at the median} 
\label{appendix:AppenFoldingForward}
\vspace{-0.30cm}
The CCDF after the first reshaping event is: 
\begin{align}
F(x,\tau^+) = \begin{cases}
1,~x\leq m(\tau),\\
\int^\infty_x dy~\rho(y,\tau^+),~x\geq m(\tau).
\end{cases}
\end{align}
In addition, $F(x,\tau) < F(x,\tau^+),~x \in (0,\infty)$. And if reshaping events are
repeated at regular intervals, then
\begin{align}
\hspace{-0.3cm}F(x,(n\tau)^+) = \begin{cases}
1,~x\leq m(n\tau),\\
\int^\infty_x dy~\rho(y,(n\tau)^+),~x\geq m(n\tau),
\end{cases}
\end{align}
with the region over which the CCDF equals unity expanding with each reshaping event. Now, at any reshape
event, the following holds true: $F(m(n\tau),n\tau) = 1/2$ and $F(m(n\tau),(n\tau)^+) = 1$. And
the jump behavior of the median implies that $F(x,n\tau) < F(x,(n\tau)^+)~\forall~x\geq m(n\tau)$.
As the random walk starts in the region $x \in [m(n\tau),\infty)$ with probability one at $t = (n\tau)^+$,
temporal evolution of the random walk in the time interval $t \in ((n\tau)^+,(n+1)\tau)$ will be such
that $F(m(n\tau),t) = c_0(n)$. Moreover, the quasi-steady state of $F(m(n\tau),\tau)$ implies that
$\exists~N$ depending on $\tau$ such that for $n\geq N$, the random
walk practically does not feel the reflecting wall for its evolution in the time interval
$t \in ((n\tau)^+,(n+1)\tau)$. This implies that
$c_0$ is almost independent of $n$ for $n\geq N$. As a result, the CCDF $F(m(n\tau),t)$ decreases
from $1$ to $c_0$ for every time interval $t \in ((n\tau)^+,(n+1)\tau)$ and $n \geq N$. This results in
the increment in the median $m((n+1)\tau)-m((n\tau)^+)$ becoming almost independent of $n$,
provided $n \geq N$.
As the argument holds for any other quartile (also percentile), the above analysis shows that for
$n \geq N$, the jump in the median is fixed, and its increment over the interval before the next reshape
is also fixed. As a result, the CCDFs for $n\geq N$ are fixed shapes, traveling at fixed speed.

\section{Forward reshaping with large $\tau$} 
\label{AppenApproxBMFwReshaping} 
\vspace{-0.2cm}Here, we derive the analytical approximation for the shape of the spatial density of forward-reshaped Brownian motion. 
While the analytical formula for the general case of $D\tau$ is infeasible, it is possible to approximate the shape of the spatial density in consecutive reshapings, when $D\tau\gtrsim 1$, i.e. for a long diffusion step. Namely, regardless of the initial condition, as long as its median is at some given $x_0$, after the first reshape, the solution is a Gaussian centered at $m_1=x_0$ with variance $\sigma_1^2=2D\tau$. Upon forward reshaping, the median moves to the right by $\sqrt{2/\pi}\sigma_1$. The variance after cutting the Gaussian distribution below the median is $\sigma_1^2(1-2/\pi)$.
As a result, even though  the resulting half-Gaussian distribution is highly skewed, after evolving it for a long time  $\tau\gtrsim 1/D$,  the resulting distribution will again assume an approximate Gaussian form. The mean or median will remain unchanged (as it was right after the reshaping and before the evolution) $m_2=m_1+\sqrt{2/\pi}\sigma_1$, while the variance will be that of the pre-evolution half-Gaussian plus the additional contribution from diffusion which is $2D\tau$, thus $\sigma_2^2=\sigma_1^2(1-2/\pi)+2D\tau$. After the $n$th step, the median will become $m_n=m_{n-1}+\sqrt{2/\pi}\sigma_{n-1}$, while the variance satisfies $\sigma_n^2=\sigma_{n-1}^2(1-2/\pi)+2D\tau$.
The asymptotic limit of large number of reshapes $n$ yields a Gaussian with mean and variance satisfying 
Eq.~\eqref{EqBMFwReshaping}, with $m_n=m((n+1)\tau)$, $m_{n-1}=m(n\tau)$ and $\sigma_n=\sigma((n+1)\tau)$. 
Importantly, the accuracy of this approximation gradually deteriorates as the skewness slowly increases with time, until it converges to a stationary value that can be found numerically.

\end{document}